# Spin polarized STM imaging of nanoscale Néel skyrmions in an SrIrO₃/SrRuO₃ Perovskite Bilayer

J. P. Corbett[1a)], K.-Y. Meng ,[1] J. J. Repicky[1], R. Garcia-Diaz[2], J. Rowland[1], A. S. Ahmed,[1] N. Takeuchi[2], J. Guerrero-Sanchez[2],  F.Y. Yang[1] and J.A. Gupta[1]

[1] *Department of Physics, the Ohio State University, Columbus, OH 43210, USA*

[2] *Centro de Nanociencias y Nanotecnología, Universidad Nacional Autónoma de México, Apartado Postal*

*14, Ensenada Baja California, Código Postal 22800, Mexico*

Spin-polarized scanning tunneling microscopy (SPSTM) was used to directly image nanoscale Néel skyrmions in a SrIrO₃ / SrRuO₃ bilayer system that are among the smallest reported to date in any system. Off-axis magnetron sputtering was used to cap epitaxial films of the oxide ferromagnet SRO with 2 unit cells of SrIrO₃, intended to provide interfacial spin orbit coupling. Atomic resolution STM imaging and tunneling spectroscopy were used to identify island-like SrIrO₃ grains and small regions of bare SrRuO₃. Isolated skyrmions were only observed in SrIrO₃-covered regions of the film and exhibited a distribution of sizes and shapes with an average diameter of 3 nm. We found that skyrmions must be fully contained within, but may be smaller than, any given SrIrO₃ region. Additionally, skyrmions were observed on SrIrO₃ islands of varying thickness without loss of SPSTM contrast, suggesting the magnetic texture lies within the SrIrO₃ island rather than the underlying ferromagnetic SrRuO₃. Density functional theory calculations suggest this could be due to a small induced magnetic moment associated with IrO layers in the SrIrO₃ film.

Here we report spin polarized STM measurements that directly confirm isolated Néel-type Skymrions in the $SrIrO_3$ / $SrRuO_3$ (SIO/SRO) bilayer system with an average diameter of 3 nm, which are among the smallest reported in any system. The topological nature of the skyrmions is confirmed by SPSTM imaging with different tip spin orientations and comparison with simulated SPSTM images. The high spatial resolution and complementary topographic and spectroscopic information of SPSTM allows us to directly correlate skyrmions in the materials with the surface morphology. Isolated skyrmions were only observed in SIO-covered regions of the film but were otherwise independent of island area and thickness. Density Functional Theory calculations suggest this may reflect an induced magnetic moment in the SIO film. As skyrmion sizes approach the characteristic length scales of their local environment, understanding how skyrmions are distributed amongst the local morphology is a pivotal step toward understanding skyrmion motion and pinning at nanometer length scales.

**RESULTS AND DISCUSSION**

STM images of the SIO/SRO bilayer sample reveal large, atomically-flat areas covered by a densely packed network of bright circular islands that we attribute to the SIO layer, with occasional regions of dark contrast that we attribute to the underlying SRO film. The line profile shown in Fig. 1b shows fluctuations in the apparent height that roughly correspond to multiples of the SIO unit cell. To better distinguish SIO and SRO, we compare tunneling spectra ($dI/dV$) taken on these regions in Fig. 1c. In regions with the darkest contrast, we observe a small dip near 0V with a gradually increasing signal on either side. This spectrum is similar to control STM measurements on the bare SRO film (c.f. Supporting Information) and recent STM studies,[1] and is consistent with the conducting density of states of SRO from DFT calculations[2]. In contrast, tunneling spectra taken on the islands show a well-developed gap. While the gap width varies in a range ~ 0.5 – 1V, it is consistently observed on all of the islands, and we attribute these to SIO regions. This assignment is consistent with thin film transport studies showing increasingly insulating behavior for films < 4 u.c. thick.[3–6] Atomic resolution imaging of the islands (inset) shows regions with a (1x1) square, 0.4 nm lattice which is consistent with SIO and there are regions which show a $(1 \times \frac{\sqrt{2}}{2})$R45 row structure running along the (1×1) diagonal which we attribute to facets of the islands.

Spatial mapping of the d$I$/d$V$ signal allowed us to find isolated skyrmions associated with a fraction of the SIO islands. Figure 2a shows an STM topographic image of two irregularly-shaped SIO islands and the corresponding dI/dV map (Fig. 2b) which shows bright contrast < 10 nm in diameter associated with the islands. To interpret this contrast as magnetic in origin, we note that the Cr tips used in our SPSTM measurements are antiferromagnetic, so that the atomic termination of the tip has a preferred spin polarization direction, even though there is no net magnetization of the tip [7]. Bright SPSTM contrast is expected for a Néel skyrmion imaged with a tip sensitive to the out-of-plane magnetization (OP tip), consistent with the experimental image (Fig. 2b). However, not all candidate magnetic textures are topologically non-trivial, and several TEM studies of centrosymmetric materials have found trivial Bloch-like magnetic textures[8–10]. |For non-centrosymmetric thin film systems with interfacial DMI, SPSTM simulations have recently elucidated the range of contrast depending on texture topology and tip orientation.[11] We formed a trivial bubble texture by joining two Néel domain walls along a line, and find that simulated SPSTM images also show bright contrast for an OP tip.

To better distinguish between these possibilities, it is important to compare images with different tip spin orientations, and this can be achieved with the bulk Cr tips in our experiments as the atomic termination changes during repeated imaging.[12] Figure 2c shows a subsequent spatial map of the same two textures under otherwise identical conditions. The switching of contrast to bright/dark lobed structures is consistent with the simulated SPSTM image for a Néel skyrmion with an in-plane (IP) tip, while a more complicated clover pattern would be expected for the trivial bubble. In a similar vein, because DMI selects a handedness for the winding of the magnetization, an applied magnetic field is expected to invert each spin within the texture and thus the overall double-lobe contrast. [12] Figure 2d shows such inversion for another, 2nm skyrmion imaged with an IP tip. Though the applied out-of-plane field of ±1 T is too small to polarize the background ferromagnetic film or affect the antiferromagnetic tip, we do observe an inversion of the lobed contrast for the skyrmion. The switching of contrast with tip orientation and magnetic field in Fig. 2 allows us to rule out electronic contributions to the images and unambiguously identify these textures as Néel skyrmions stabilized by interfacial DMI.

To understand the role of the SIO islands in providing the necessary DMI, we show in Figures 3(a-b) a skyrmion localized to one SIO island within the field of view and the corresponding SPSTM map. We overlay the SPSTM signal onto the topograph in Fig. 3c, to show that while the Skymrion is associated with a particular SIO island, it's spatial extent doesn't exactly trace the island's topography. To examine this behavior more quantitatively, we compiled a statistical analysis comparing skyrmion sizes and their host island areas as shown in Fig. 3d. We find that at the lower end (< 10 nm2), skyrmions are limited by the size of the host island (orange line in Fig. 3d), while the skyrmion size is independent of island area for larger islands. Since the all data points fall on the right-hand side of the orange line, the skyrmions are never larger than the associated SIO island. A histogram of skyrmion area is plotted in Figure 3(e) and gives an average skyrmion area of $8.0 \pm 5.0$ nm$^2$ and a mode of 6.0 nm$^2$. The distribution is slightly positively skewed, reflecting the physical limit imposed by the atomic structure at the lower end of the distribution.

Since bulk SIO is paramagnetic, we expected that the SIO islands provide the large spin-orbit coupling needed for interfacial DMI and that the skyrmions are hosted in the underlying ferromagnetic SRO layer. Surprisingly, we find that additional analysis of the skyrmions with respect to SIO island heights suggests that this may not be the case. If SIO were simply a paramagnetic layer, then we would expect the SPSTM signal to decay exponentially with island thickness, as the STM tip is further removed from the SRO layer. Though SIO is insulating, our choice of bias voltage was well outside the gap region, so that tunneling electrons should preferentially be probing the surface density of states. We would thus expect to only find skyrmions on thinner SIO islands. This is not observed experimentally, as Figure 4a shows that skyrmions can be found on SIO islands of various heights ranging from 0.5 – 5.0 u.c.. The distribution peaks at 2-3 u.c. height as that is the most likely island height due to the nominal 2 u.c. film thickness. We would also expect an exponential decrease in the SPSTM contrast on thicker SIO islands, while we instead observe that the SPSTM signal is roughly constant with island thickness (Fig. 4b).

Several studies in heterostructure samples suggest that SIO can acquire an induced magnetic moment from neighboring magnetic layers, of order (~0.04-0.06 $\mu_B$/Ir[13,14]), which may explain how we can observe the skyrmions. To explore this possibility, we performed spin-polarized DFT calculations of the SIO/SRO bilayer system. In the 10 u.c. SRO film, we find that the magnetic moment ($\mu B$) per atomic layer primarily lies within the RuO planes (1.5 $\mu_B$/Ru), with a much smaller induced moment on the SrO planes (~0.3 $\mu_B$/Sr). Total energy calculations suggest that SrO is the preferred terminating layer, which is in agreement with prior STEM on these samples. We find a similar induced moment on the IrO planes in the overlying SIO film (0.1-0.2 $\mu_B$/Ir), also in agreement with previous DFT [15–17]. While this moment persists for all IrO planes in the SIO film, the induced moment in the SrO planes in SIO falls quickly to zero away from the SRO interface. DFT total-energy calculations with varying chemical potential suggest that the magnetic IrO termination is preferred over most of the Sr/Ru phase space, but SrO terminations may also be possible. This raises the possibility of co-existing magnetic (IrO) and nonmagnetic (SrO) terminating surfaces in our SIO film, which could explain why we observe Skrymions on only a fraction of the islands. Our histograms of island height were broken up into ½ u.c. bins to account for this possibility (c.f. Supporting Information), but we note that the small island size and curved shape adds a comparable ~½ u.c. uncertainty to the height measurements.

In conclusion, we report the discovery of few-nm Néel-type skyrmions in SIO/SRO bilayers that are among the smallest known in any system to date. The skyrmions are directly correlated with SIO-covered regions, but except in the case of atomically small islands, exhibit a size that reflects the competition of DMI and ferromagnetic exchange rather than the island size itself. The accessibility of these skyrmions to surface-sensitive SPSTM suggests that they may reflect an induced magnetic moment in the SIO, which is consistent with prior studies. This study probes a novel regime where the Skymrion and film morphology have similar length scales, which will be important for harnessing nanoscale skyrmions for high density magnetic memories. The broad tunability of oxide systems is particularly promising as a platform for such devices.

**METHODS**

**Sample growth:** Epitaxial growth of the perovskites bilayer ultra-thin films was achieved using a custom-built off-axis dc sputtering system. The commercial $SrTiO_3$(001) substrates were cleaned with a buffered-HF solution for 30 s and subsequently annealed in ultra-high vacuum (UHV) for 1 hour at 950 °C. The substrate was then cooled to the growth temperature of 420 °C for SRO and SIO depositions. An $O_2$/Ar gas mixture was used for sputtering with a total pressure of 8 mTorr (7 mTorr) with $O_2$ partial pressures being 96 µTorr (35 µTorr) for SRO (SIO) deposition respectively. Further details of the growth, and characterization of the film and epitaxial interface quality by x-ray diffraction and cross-sectional scanning transmission electron microscopy were reported previously [18].

**STM imaging:** Following growth, the samples were transferred *ex-situ* to the UHV load lock of a Createc LT-STM system. The samples were annealed under UHV conditions to 120 °C for 1 hour to remove adsorbed air containments. The sample was then cooled to 100 K and a 0.4 T out-of-plane magnetic field was applied to saturate the magnetization of the as-grown film. The magnetic field was removed and the sample then transferred to the cold STM at 5K. All SP-STM measurements were performed at 5K using bulk Cr tips which were electrochemically etched and cleaned via cycles of $Ar^+$ sputtering. Tunneling spectra were obtained by disabling the STM feedback loop, adding a modulation voltage (50 mV$_{rms}$, 1500 Hz) to the DC sample bias, and using a lock in amplifier to measure the corresponding modulation in tunneling current d$I$/d$V$ as the DC bias was swept. Simultaneous topography and SPSTM images of the dI/dV signal were acquired at fixed voltage while the STM feedback loop was engaged. Image processing and analysis was performed using Gwyddion for STM data[19].

**Density Functional Theory:** Our spin-polarized total energy calculations are carried out within the DFT framework as developed in the Vienna Ab initio Simulation Package (VASP)[20]. To sample the electron-electron interactions, the generalized gradient approximation as stated by Perdew-Burke-Ernzerhof has been used[21]. Projector augmented-waves were used to treat the core electrons [22]. Electronic states were expanded in plane waves with an energy cutoff of 550 eV. A gamma centered K-points grid of 6x6x1 was used to evaluate the Brillouin zone integration. To evaluate the electronic properties, a denser K-points mesh of 12x12x1 was used.

In order to model the bilayer system, we first determined the most stable lattice parameters for the SrRuO3, and SrIrO3 perovskites in their cubic phase. After structural optimizations, the obtained lattice parameters were 3.97 Å and 4.02 Å, respectively. The $SrRuO_3(001)$ surfaces were constructed by using the supercell method. We considered the two possible surface terminations emerge, SrO and RuO (we have used slab thickness of 8 and 7 layers, respectively). A vacuum space of 15 Å was used to preclude surface self-interactions on both models. Once optimized, we proceed to place on top two $SrIrO_3$ unit cells, with its lattice parameter set to match the SrRuO3 substrate. Two possible surface terminations again emerge, Sr-O and Ir-O and we utilize 12 and 13 layers respectively with the same 15 Å vacuum space.

## Acknowledgements

We acknowledge primary support from the Defense Advanced Research Projects Agency (Grant No. D18AP00008). We thank DGAPA-UNAM projects IA100920 and IN101019, and CONACYT grants A1-S-9070 for partial financial support. Calculations were performed in the DGCTIC-UNAM supercomputing center project LANCAD-UNAM-DGTIC-368. The authors would like to thank E. Murillo and A. Rodriguez Guerrero for their technical assistance and useful discussions.

Author Contributions

Competing Interests statement

**REFERENCES**

[1] H.G. Lee, L. Wang, L. Si, X. He, D.G. Porter, J.R. Kim, E.K. Ko, J. Kim, S.M. Park, B. Kim, A.T.S. Wee, A. Bombardi, Z. Zhong, and T.W. Noh, Advanced Materials **32**, 1905815 (2020).

[2] G. Koster, L. Klein, W. Siemons, G. Rijnders, J.S. Dodge, C.-B. Eom, D.H.A. Blank, and M.R. Beasley, Reviews of Modern Physics **84**, 253 (2012).

[3] F.-X. Wu, J. Zhou, L.Y. Zhang, Y.B. Chen, S.-T. Zhang, Z.-B. Gu, S.-H. Yao, and Y.-F. Chen, Journal of Physics: Condensed Matter **25**, 125604 (2013).

[4] A. Biswas, K.-S. Kim, and Y.H. Jeong, Journal of Applied Physics **116**, 213704 (2014).

[5] P. Schütz, D. Di Sante, L. Dudy, J. Gabel, M. Stübinger, M. Kamp, Y. Huang, M. Capone, M.-A. Husanu, V.N. Strocov, G. Sangiovanni, M. Sing, and R. Claessen, Physical Review Letters **119**, (2017).

[6] D.J. Groenendijk, C. Autieri, J. Girovsky, M.C. Martinez-Velarte, N. Manca, G. Mattoni, A.M.R.V.L. Monteiro, N. Gauquelin, J. Verbeeck, A.F. Otte, M. Gabay, S. Picozzi, and A.D. Caviglia, Phys. Rev. Lett. **119**, 256403 (2017).

[7] R. Wiesendanger, Reviews of Modern Physics **81**, 1495 (2009).

[8] Z. Hou, W. Ren, B. Ding, G. Xu, Y. Wang, B. Yang, Q. Zhang, Y. Zhang, E. Liu, F. Xu, W. Wang, G. Wu, X. Zhang, B. Shen, and Z. Zhang, Advanced Materials **29**, 1701144 (2017).

[9] X. Yu, Y. Tokunaga, Y. Taguchi, and Y. Tokura, Advanced Materials **29**, 1603958 (2017).

[10] X. Yu, M. Mostovoy, Y. Tokunaga, W. Zhang, K. Kimoto, Y. Matsui, Y. Kaneko, N. Nagaosa, and Y. Tokura, Proceedings of the National Academy of Sciences **109**, 8856 (2012).

[11] K. Palotás, L. Rózsa, E. Simon, L. Udvardi, and L. Szunyogh, Physical Review B **96**, (2017).

[12] N. Romming, A. Kubetzka, C. Hanneken, K. von Bergmann, and R. Wiesendanger, Physical Review Letters **114**, (2015).

[13] J. Nichols, X. Gao, S. Lee, T.L. Meyer, J.W. Freeland, V. Lauter, D. Yi, J. Liu, D. Haskel, J.R. Petrie, E.-J. Guo, A. Herklotz, D. Lee, T.Z. Ward, G. Eres, M.R. Fitzsimmons, and H.N. Lee, Nature Communications **7**, (2016).

[14] D. Yi, J. Liu, S.-L. Hsu, L. Zhang, Y. Choi, J.-W. Kim, Z. Chen, J.D. Clarkson, C.R. Serrao, E. Arenholz, P.J. Ryan, H. Xu, R.J. Birgeneau, and R. Ramesh, Proceedings of the National Academy of Sciences **113**, 6397 (2016).

[15] W. Fan and S. Yunoki, Journal of Physics: Conference Series **592**, 012139 (2015).

[16] K.-H. Kim, H.-S. Kim, and M.J. Han, Journal of Physics: Condensed Matter **26**, 185501 (2014).

[17] T.R. Dasa, L. Hao, J. Yang, J. Liu, and H. Xu, Materials Today Physics **4**, 43 (2018).

[18] K.-Y. Meng, A.S. Ahmed, M. Baćani, A.-O. Mandru, X. Zhao, N. Bagués, B.D. Esser, J. Flores, D.W. McComb, H.J. Hug, and F. Yang, Nano Letters **19**, 3169 (2019).

[19] D. Nečas and P. Klapetek, Open Physics **10**, (2012).

[20] G. Kresse and J. Furthmüller, Physical Review B **54**, 11169 (1996).

[21] J.P. Perdew, K. Burke, and M. Ernzerhof, Physical Review Letters **77**, 3865 (1996).

[22] P.E. Blöchl, Physical Review B **50**, 17953 (1994).

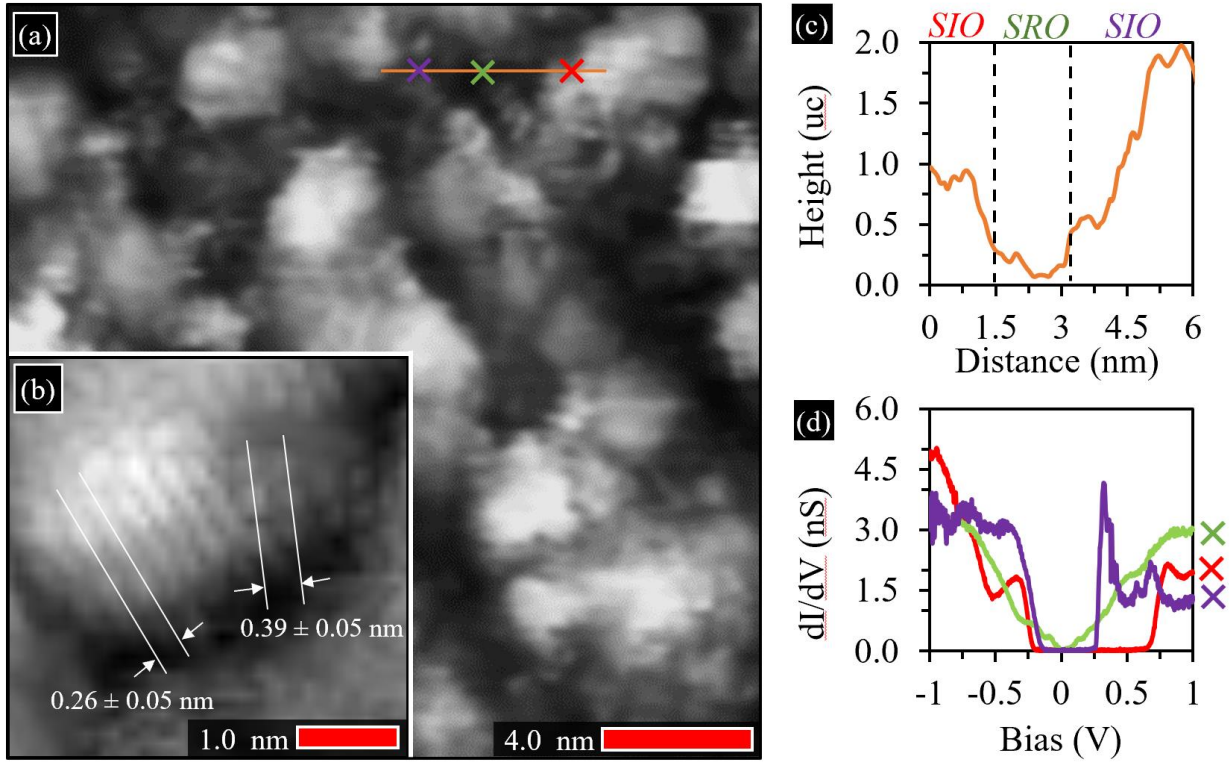

**Figure 1. Morphology of the terminating SIO layer.** (a) STM topographic image showing small bright grains attributed to SIO. (b) Atomic resolution image island showing lattice fringes consistent with SIO. (c) Line profile as indicated by the line in (a), showing height variations of ~ 1 SIO unit cell. (d) dI/dV spectroscopy corresponding to the three points marked in (a). Gap-like features are observed on SIO regions (red, purple curves), while more metallic conduction is attributed to regions where the underlying SRO is exposed (green curve).

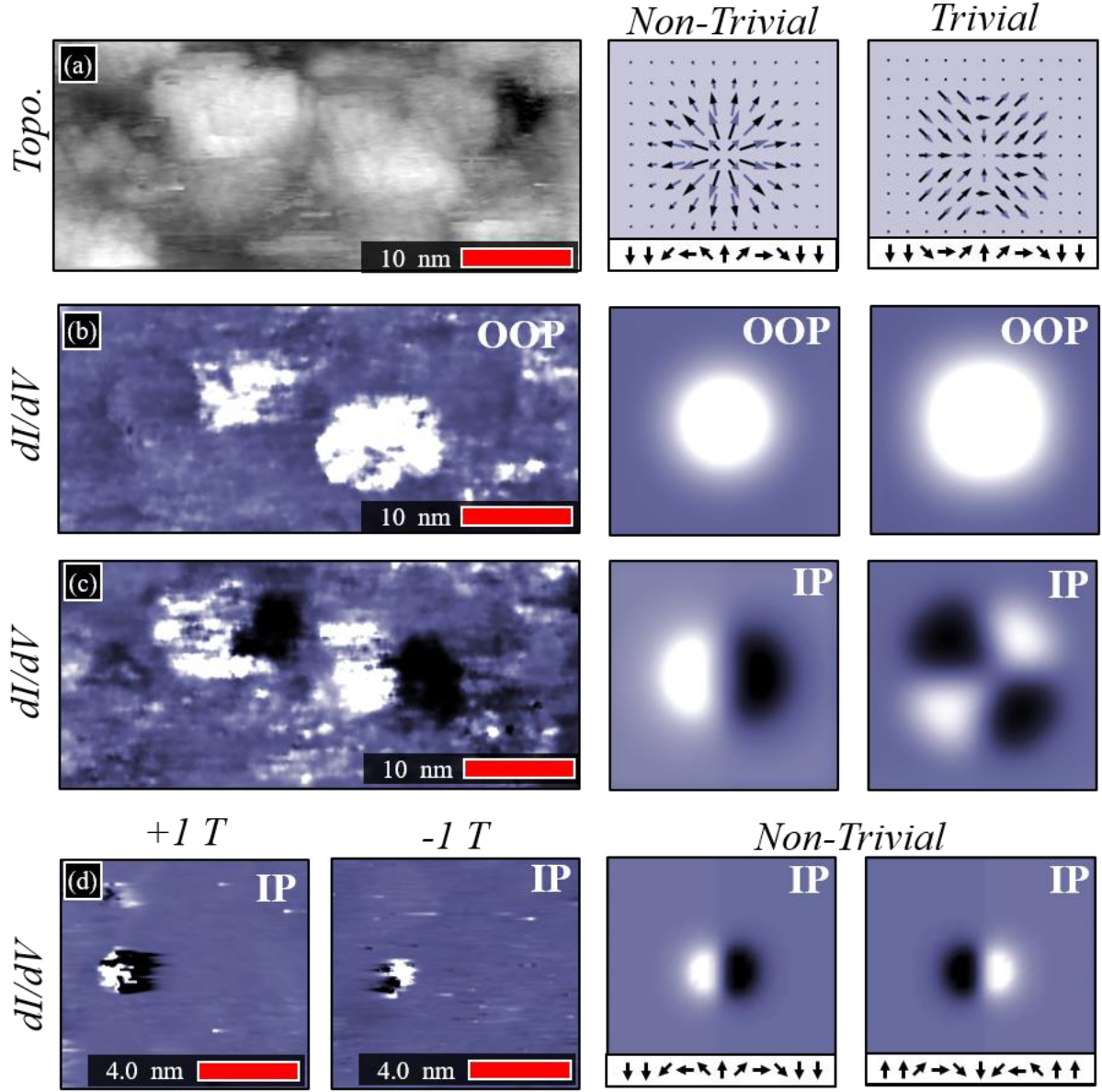

**Figure 2. Comparison of experimental and simulated SPSTM images of Neél skyrmions.** (a) Topographic STM image showing SIO grains with some lattice fringe contrast. (b) Simultaneous dI/dV image acquired with (a), showing bright contrast attributed to two skyrmions localized to SIO grains and imaged with a tip sensitive to out-of-plane (OOP) magnetic contrast. (c) dI/dV image of the same area, taken with a tip sensitive to in-plane (IP) magnetic contrast. (d) dI/dV image of a different skyrmion, showing reversal in the lobed contrast with +-1 T magnetic field applied out of the plane. Corresponding dI/dV simulations of a Neél-type skyrmion, and a trivial Neél-like bubble are shown in the right panels.

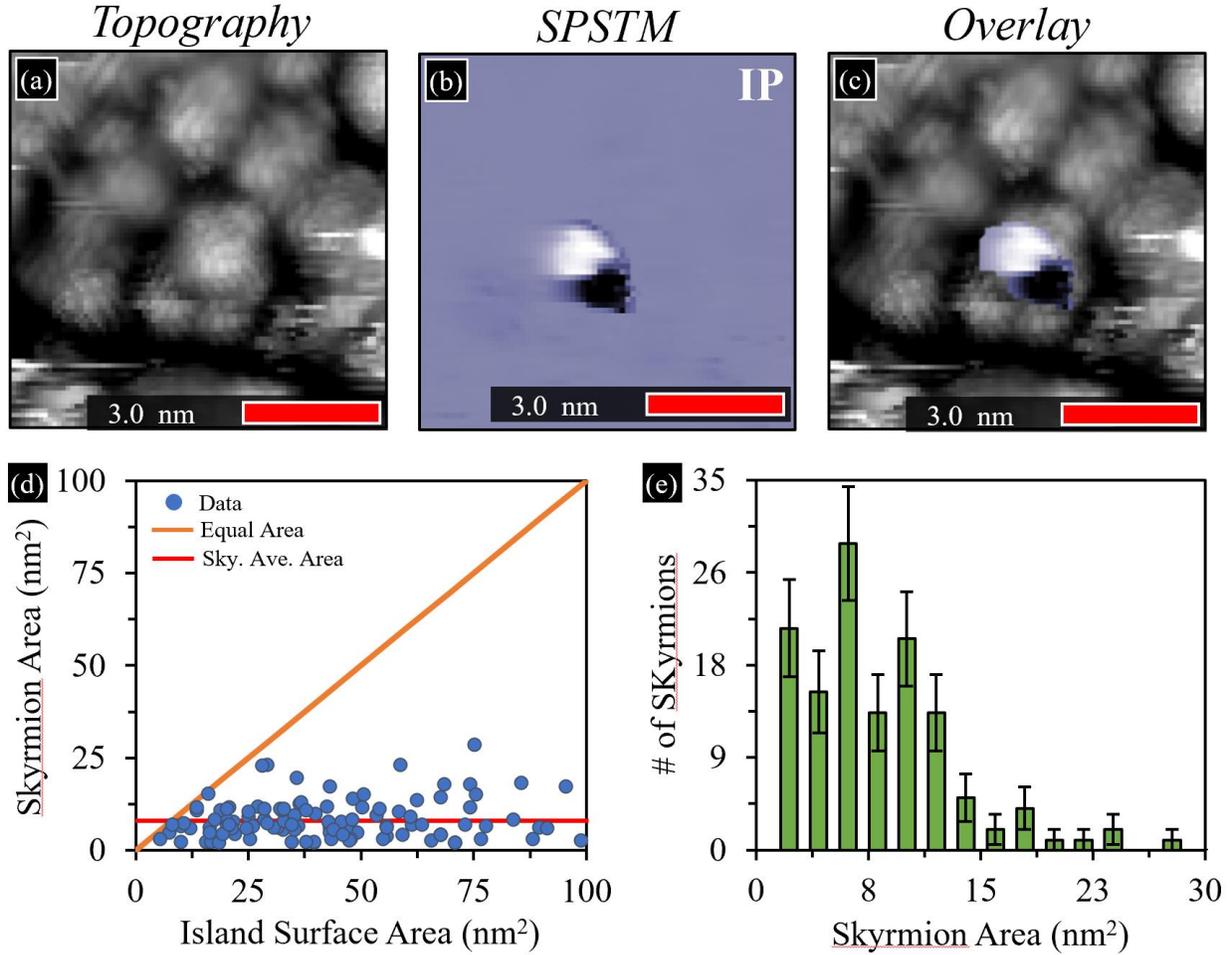

**Figure 3. Correlation of nanoscale Neél skyrmions with SIO grains.** (a) STM topographic image of few nm SIO grains. (b) Simultaneous dI/dV image, showing lobed contrast of a < 2 nm Neél skyrmion imaged with an IP tip. (c) overlay of the topographic and SPSTM images, showing the skyrmion is localized to an individual SIO island. (d) plot comparing skyrmion areas from dI/dV images, and corresponding SIO island areas. The orange line indicates equal areas; all data points fall below this line, indicating that skyrmion area is independent of SIO island size, except in the very smallest islands. The red line indicates the average skyrmion area. (e) Histogram of skyrmion sizes, with a positively skewed distribution with an average area of 8 nm².
.

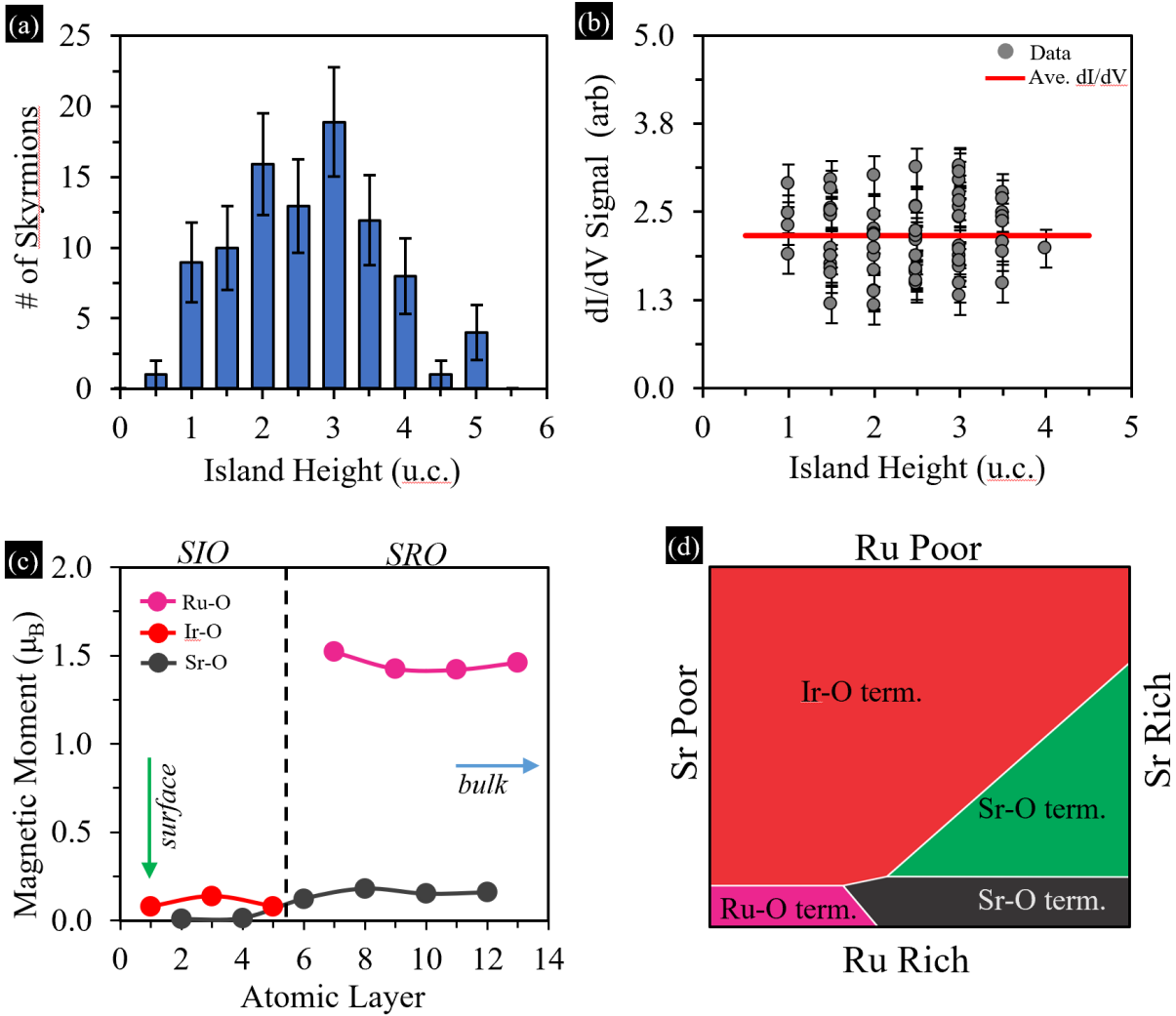

**Figure 4**. (a) Histogram of skyrmions and SIO island heights, showing a peak near the nominal 2 u.c. film thickness. (b) plot of skyrmion dI/dV contrast taken with an OOP tip versus SIO island height. Because SPSTM is surface sensitive, the absence of a pronounced dependence on separation from the ferromagnetic SRO layer is suggestive of skyrmions being hosted in the SIO itself. (c) DFT calculations of the computed magnetic moment per atomic layer for an IrO-terminated SIO/SRO heterostructure. A small induced moment is indicated on IrO layers in the SIO film. (d) DFT-calculated phase diagram of the energetically favored surface termination under varying Sr, Ru growth conditions.